\documentclass[a4paper]{jpconf}
\usepackage{graphicx}
\usepackage{xspace}
\usepackage{iopams}
\usepackage{acronym}
\usepackage{hyperref}

\newcommand{\Fstat}{$\mathcal{F}$-statistic\xspace}
\newcommand{\win}{\varpi}
\newcommand{\Dop}{\lambda} 
\newcommand{\Amp}{\mathcal{A}} 
\newcommand{\TP}{\mathcal{T}} 
\newcommand{\Btrans}{B_{\mathrm{tS}/\mathrm{G}}} 
\newcommand{\fgw}{f} 

\begin{document}
\title{Search setup for long-duration transient gravitational waves from glitching pulsars during LIGO--Virgo third observing run}

\author{Luana M. Modafferi$^1$, Joan Moragues$^1$ and David Keitel$^1$ for the LIGO Scientific Collaboration, the Virgo Collaboration and the KAGRA Collaboration}

\address{$^1$ Departament de Física, Institut d’Aplicacions Computacionals i de Codi Comunitari (IAC3),
Universitat de les Illes Balears, and Institut d’Estudis Espacials de Catalunya (IEEC),
Crta. Valldemossa km 7.5, E-07122 Palma, Spain}

\ead{luana.modafferi@ligo.org}

\begin{abstract}
Pulsars are spinning neutron stars which emit an electromagnetic beam. We expect pulsars to slowly decrease their rotational frequency. However, sudden increases of the rotational frequency have been observed from different pulsars. These events are called “glitches” and they are followed by a relaxation phase with timescales from days to months. Gravitational wave (GW) emission may follow these peculiar events. We give an overview of the setup for an analysis of GW data from the Advanced LIGO and Virgo third observing run (O3)~\cite{LIGOScientific:2021quq} searching for transient GW signals lasting hours to months after glitches in known pulsars. The search method consists of placing a template grid in frequency--spindown space with fixed grid spacings. Then, for each point we compute the transient \Fstat which is maximized over a set of transient parameters like the duration and start time of the potential signals. A threshold on the detection statistic is then set, and we search for peaks over the parameter space for each candidate.
\end{abstract}

\section{Introduction}
Continuous gravitational waves (CWs) are long-lasting, quasi-monochromatic signals which are expected to be emitted by systems with slowly evolving frequency, such as isolated neutron stars (NSs). 
NSs represent a promising laboratory of nuclear physics at high densities. Spinning NSs are expected to emit CWs when a quadrupolar deformation of the star is present, and they can be the key to answering many open questions concerning the evolution and the equation of state of the star~\cite{lasky_2015}.
Despite numerous targeted, directed and all-sky searches, CWs from pulsars are yet to be detected as these signals are very faint. 

It has been observed in the electromagnetic emission of spinning NSs that some sources exhibit spin-up events, or glitches~\cite{2017A&A...608A.131F}, which consist of the sudden rise of the rotational frequency of the star, followed by a relaxation phase with a timescale between days to months.
These peculiar events may trigger the emission of long-duration transient-CWs~\cite{PhysRevD.84.023007}, which are modelled similarly to persistent CWs, but additionally modulated by a window function $\win(t;t_0,\tau)$ which effectively limits the duration of the CW:
\begin{equation}
 \label{eq:htCW}
 h(t,\Dop,\Amp,\TP) = \win(t,t_0,\tau)\,h(t,\Dop,\Amp) \,,
\end{equation}
where $ h(t,\Dop,\Amp)$ corresponds to a CW~\cite{PhysRevD.58.063001} with amplitude parameters $\Amp$, and phase parameters $\Dop$, and the transient parameters $\TP$ consist of the window shape, signal start time $t_0$, and a duration parameter $\tau$.

There are different astrophysical motivations justifying the existence of these signals, based for example on the two-fluid model~\cite{10.1046/j.1365-8711.2000.03415.x} which describes the physics behind pulsars glitches. The transfer of angular momentum from the inner layers to the outer layers of the neutron star can induce changes in the quadrupole moment of the star and thus the emission of GWs during the aftermath of a glitch.
Previous attempts at finding these types of signals from glitching pulsars have been made in the past. The first search looked for short-duration transients on occasion of Vela’s glitch from 2006~\cite{PhysRevD.83.042001}. The first search for long-duration transients (from hours up to 120 days) was made on the O2 open data, with both Vela and Crab having glitched during that time~\cite{Keitel:2019zhb}. For both searches, no detection has been made.

Here we will focus on the methods and the search setup for long-duration transient-CWs from pulsars which have glitched during the O3 observing run of Advanced LIGO~\cite{TheLIGOScientific:2014jea} and Advanced Virgo~\cite{2015CQGra..32b4001A}. O3 is separated in two different sections by a commissioning break: O3a which ran from April 1, 2019, 15:00 UTC until October 1, 2019 15:00 UTC and O3b which ran from November 1 2019, 15:00 UTC, to March 27 2020, 17:00 UTC~\cite{aLIGO:2020wna}. 

\section{Target selection}
We target those pulsars that have glitched during the O3 observing run and that have a GW frequency $\fgw$ (assumed to be twice the rotation frequency of the pulsar) of at least 10 Hz, below which the sensitivities of LIGO and Virgo decrease significantly.
We select targets from the ATNF pulsar catalog~\cite{2005AJ....129.1993M} and the ATNF and Jodrell glitch catalogs~\cite{ATNFGlitchDatabase, JodrellGlitchCatalogue}. Additional glitches were directly selected from ephemerides provided by NICER~\cite{Ho:2020oel,Ho:2020vxt} and UTMOST radio observations~\cite{Lower:2019awx}.

Our list of targets includes: J0534$+$2200, i.e. the Crab pulsar, J0537--6910, J0908--4913, J1105--6107, J1813--1748 and J1826--1334. All targets have glitched once during O3, except for J0537--6910, also known as ``the Big Glitcher"~\cite{Middleditch:2006ky} for its intense glitch activity, from which four glitches were observed during the time of interest~\cite{Ho:2020vxt}.

\section{Search method}
The method we used has been suggested in~\cite{PhysRevD.84.023007} and first applied in~\cite{Keitel:2019zhb}. The transient-CW signal is modelled as in Eq.~\ref{eq:htCW}. We define a likelihood ratio between the signal and Gaussian noise hypotheses, which we call the \Fstat matrix $\mathcal{F}_{mn}=\mathcal{F}(t_{0m},\tau_n)$ where $t_{0m},\tau_n$ refer to different signal start time and signal duration choices.
We perform the search using the \texttt{lalapps\_ComputeFstatistic\_v2} program available in LALSuite~\cite{LALSuite, Prix:2015cfs}.
The program operates in such a way that it places a grid in frequency and spindown parameters $\fgw^{(k)}$, where the maximum $k$ depends on the availability of the spindowns in the ephemerides, at a fixed sky location by using the option \texttt{gridType=8}, which corresponds to the metric from~\cite{2007PhRvD..75b3004P} and a template placement algorithm from~\cite{Wette_2014}.
We fix the metric mismatch, which determines how densely templates are placed, to 0.2. 
At each point of the grid, the program computes the per-SFT \Fstat atoms, which are then partially summed when looping over the transient parameters $t_0$ and $\tau$.  
One can then either maximize $\mathcal{F}_{mn}$ over the transient parameters $t_0$ and $\tau$ obtaining $\max{\mathcal{F}_{mn}}$~\cite{Keitel:2019zhb} or marginalize $\mathcal{F}_{mn}$ over uniform priors on $t_0$ and $\tau$ resulting in the Bayes factor $\Btrans$ for transient-CW signals against Gaussian noise.
We decide to use $\Btrans$ as our detection statistic, as it produces cleaner background distributions of the data. We use a simple rectangular window function, as it is the least computationally expensive, which effectively ``turns on" the transient GW at $t_0$ and turns it off at $t_0+\tau$. It has been shown~\cite{PhysRevD.84.023007, Keitel:2019zhb} that the loss of signal-to-noise ratio (SNR) when using a rectangular window instead of a more realistic exponentially decaying window is acceptable.

\section{Search setup}
\subsection{GW data}
We use 1800\,s long Tukey-windowed Short Fourier Transforms (SFTs) for all three detectors. These data products are derived from the calibrated, cleaned and gated time-domain data~\cite{LIGO:2021ppb,VIRGO:2021umk,Sun:2020wke,Sun:2021qcg,T2000384}. In addition, we clean bins of the SFTs affected by known instrumental line artifacts~\cite{LIGO:2021ppb,LSC:2018vzm,linelist01,virgo3lines} by replacing them with Gaussian noise matching the noise level in the surrounding range.

\subsection{Time domain setup}
In this section we explain the choices we make for the transient parameters $t_0$ and $\tau$.

For the start time of the transient $t_0$ our general rule is to search over a range of different start times in the interval:
\begin{equation}
    \label{eq:t0}
    t_0 \in [T_{\mathrm{gl}} - \max(\delta T_{\mathrm{gl}}, 1\,\mathrm{day}), T_{\mathrm{gl}} + \max(\delta T_{\mathrm{gl}}, 1\,\mathrm{day})],
\end{equation}
where $T_{\mathrm{gl}}$ and $\delta T_{\mathrm{gl}}$ are the time at which the glitch was detected from the EM observations and its uncertainty respectively.
However, depending on the available SFTs, there can be exceptions to this rule, when for instance the interval of start times does not overlap with the available SFTs.
When there is no available data  in the $t_0$ interval, we shift the start time to the first available SFT timestamp and set the interval band to 0: according to our choice above, 
the  $t_0$ band should be centered at the $T_{\mathrm{gl}}$, so if there is no data in the $t_0$ interval of interest, we do not search around that value.

We then decide to search for transients with duration $\tau$ in the interval: $\tau \in [3600 $ s, 120 days$]$.
However, as before, there can be exceptions.
Taking J0537--6910 as example, this pulsar suffered 4 glitches in the O3 time range, with an inter-glitch interval as small as 61 days and as big as 170 days. For those glitches with an interglitch period smaller than the 
chosen 120 days, we simply cut the duration interval to the timestamp of the following glitch, $T_{\mathrm{gl}}^{n+1}$. We perform the same operation when the observing run ends before the 120 days from the glitch. Therefore one can summarize by taking:
\begin{equation}
    \label{eq:tau}
    \tau \in [3600\textnormal{ s}, T_{\mathrm{end}}-\min(t_0)] \textnormal{ or } \tau \in [3600\textnormal{ s}, T_{\mathrm{gl}}^{n+1}-\min(t_0)].
\end{equation}

\subsection{Frequency domain setup}
There are different methods to search for GWs from pulsars, and in particular this work uses the narrow-band approach, meaning that the sky position, the frequency and spindowns (the phase evolution parameters) are all known parameters. But unlike the targeted searches in which the GW signal is phase-locked to the EM ephemerides, we here allow some uncertainty
in the frequency and spindowns around their electromagnetically-determined values.

We relax the assumption of the GW frequency to be exactly twice the rotational frequency because either the electromagnetic ephemerides might not always be accurate, or there may even be a physical reason. Such a mismatch between the true value 
and the observed one in the radio or X-ray counterpart may prevent the GW signal from being detected~\cite{PhysRevD.99.122002}.
The small mismatch can be parametrized as follows:
\begin{equation}
    \label{eq:narrow_band_delta}
    \Delta \fgw = \fgw(1+\delta),
\end{equation}
where we use $\delta = 10^{-3}$ for this search.

The frequency and spindown bands we choose are essentially:
\begin{equation}
    \label{eq:sd_bands}
    \Delta \fgw^{(k)} = \max(2\times \delta \fgw^{(k)}, 10^{-3}\times \fgw^{(k)}, \textnormal{ glitch step})
\end{equation}
where $\fgw^{(k)}$ is the GW frequency or derivative extrapolated at the reference time and $\delta \fgw^{(k)}$ is its uncertainty including the propagation of the glitch uncertainties. The maximum value of $k$ depends on the ephemerides and is at most $k=3$. 
We set the reference time for the search grid to the first available SFT timestamp.  
The $\textnormal{glitch step}$ is the difference between the spindown before and after the glitch, including the uncertainties.

\subsection{Choice of threshold}
In order to separate the candidates to follow up from noise, we set a threshold on $\Btrans$. The proper way to establish such a threshold
would be to use the distribution of the expected maximum of $\Btrans$. However, this is not known analytically. We use the practical solution suggested in~\cite{Tenorio:2021wad},
which applies extreme value theory to empirically estimate this distribution from the search results themselves.

\subsection{Upper limits}
If there are no detections for a given target, we compute upper limits on the GW strain of the transient-CWs. We mainly follow the procedure used in~\cite{Keitel:2019zhb}.

We perform software injections of simulated signals into the same data sets as the ones used for the original searches. We inject into uncleaned data, so to ensure we are treating the software injections the same as if they had been present in the original data, and not overestimate our detection probability by falsely counting injections in the cleaned bands as detectable. For a certain set of durations $\tau$ we perform several injections, with several steps in strain $h_0$, uniformly distributed over the search ranges of frequency, spindown and start time $t_0$ and randomized over the other amplitude parameters.
We then perform a search over a small band on these injected signals, and consider a signal as recovered if its $\Btrans$ is above the detection threshold.
For each $\tau$ the result is an efficiency curve of the probability of detection $p_{\mathrm{det}}$ against injected $h_0$. This can be fit with a sigmoid function, and we find the value $h_0^{95\%}$, at which $p_{\mathrm{det}} = 95\%$.
This is the estimate of the required scale of the strain at which $95\%$ of the signals are recovered above the threshold on $\Btrans$. We repeat these steps for the rest of the durations 
and collect $h_0^{95\%}$ as a function of $\tau$. 

\section{Conclusions}
In this contribution we have summarized the search setup and method of a search for long-duration transient-CWs from pulsars that have glitched during Advanced LIGO and Advanced Virgo's third observing run. The results are reported in an LVK paper~\cite{LIGOScientific:2021quq} which includes also the narrow-band searches for persistent CWs from a larger set of known pulsars.

\ack
L.M.M. is supported by the Universitat de les Illes Balears.
D.K. is supported by the Spanish Ministry of Science and Innovation (ref. BEAGAL 18/00148) and cofinanced by the Universitat de les Illes Balears.
This work was supported by the European Union FEDER funds; the Spanish
Ministry of Science and Innovation and the Spanish Agencia Estatal de Investigación grants
PID2019-106416GB-I00/AEI/MCIN/10.13039/501100011033, RED2018-102661-T, RED2018-102573-E;
the Comunitat Autonoma de les Illes Balears through the Direcció General de Política Universitaria i
Recerca with funds from the Tourist Stay Tax Law ITS 2017-006 (PRD2018/24); the Conselleria de Fons Europeus, Universitat i Cultura del Govern de les Illes Balears; the Generalitat
Valenciana (PROMETEO/2019/071); and EU COST Actions CA18108, CA17137, CA16214, and CA16104.
LVK acknowledgements may be found in \texttt{https://dcc.ligo.org/P2100218} .

\section*{References}
\bibliography{iopart-num}
\end{document}